\documentstyle[12pt]{article}
\begin{document}
\begin{large}
\leftline{\hskip 12 truecm IJS.TP.00/1} 

\vspace{5mm}

\centerline{\bf The MAJORANA PARTICLES and the MAJORANA SEA}

\end{large}

\vspace{0.8cm}
\centerline{\rm NORMA MANKO\v C BOR\v STNIK\footnote[1]{The
invited talk, presented on International Workshop on Lorentz
group, CPT and Neutrinos, Zacatecas, 23-26 June, 1999.}}
\centerline{\it Dept. of Physics, University of
Ljubljana, Jadranska 19, and }
\centerline{\it J. Stefan Institute, Jamova 39,
Ljubljana, 1111, and}
\centerline{\it Primorska Institute for Natural
 Sciences and Technology, 6000 Koper, Slovenia}
\centerline{\it E-mail: norma.s.mankoc@ijs.si }

\vspace{2mm}
\centerline{\rm and}
\centerline{\rm HOLGER BECH NIELSEN}
\centerline{\it Dept. of Physics,  Niels Bohr Institute,}
\centerline{\it Blegdamsvej 17, Copenhagen, DK-2100 and}
\centerline{\it Theoretical Physics Division, CERN, CH-1211
Geneva 23}
\centerline{\it E-mail: hbech@nbvims.nbi.dk}

\vspace{2mm}
\centerline{\rm and}
\centerline{\rm COLIN D. FROGGATT}
\centerline{\it Dept. of Physics and Astronomy, Glasgow University,}
\centerline{\it Glasgow G12 8QQ, Scotland, UK}
%\centerline{\it E-mail: hbech@nbvims.nbi.dk}

\abstract{ Can one make a Majorana field theory for fermions
starting from 
the zero mass Weyl theory, then adding a mass term as an interaction?  
The answer to this question is: yes we can. We can proceed 
similarly to the case of the Dirac massive field
theory\cite{jda}. In both 
cases one can start from the zero mass Weyl theory and then 
add a mass 
term as an interacting term of massless particles with a
constant ( external )
field. In both cases the interaction gives rise to a field
theory for a free
massive fermion field. We  present the procedure for the
creation of a mass term in the case of the Dirac and the
Majorana field and we  look for a massive field as
a superposition of massless fields.}

\section{  Introduction\label{int}}

\noindent
A Majorana fermion, being its own antifermion, is an
unusual particle. If charges
should be conserved, the Majorana particle should carry no
charge. Among bosons the photon is its own antiparticle, that is
a "Majorana boson". Among fermions, neutrinos are candidates for
Majorana particles.

\noindent
If you count neutrinos as belonging to the Standard model, the
right handed neutrinos appear with no 
charge and can accordingly interact only through the
gravitational force. Going beyond the Standard model, 
additional quantum numbers, connected with additional gauge
fields, can be designed for fermions, and accordingly also for
 right handed neutrinos. If one 
starts, for example, with the group SO(1,13) to describe all the
internal degrees of freedom (see refs.(\cite{nmb}, \cite{proc}) of
fermions and bosons in an unique way or
with the group SO(10) to describe uniquely only charge degrees of
freedom, one additional quantum number appears, connected with a
gauge field, which is nonzero for left and right handed quarks
and leptons, except for charged leptons, and enables the right
handed neutrino to interact.

\noindent
In this talk we shall only present the parallelism between the
change of the Dirac (Weyl) sea of massless fermions, for two
kinds of interactions, one leading to massive Dirac fermions,
the other to massive Majorana fermions. We also shall present
for both kinds of the sea the excitations of the sea, representing the 
physicsl fermions.
This presentation may help to better understand the nature of
the Dirac and the Majorana particles.
In this talk we pay attention on only spin degrees of freedom.
The dimension of the ordinary space-time is four.

\section{ The zero mass Weyl field theory \label{zm}}

\noindent
We  start with massless fermions, described by the Weyl
massless fields. We pay 
attention on a momentum $(p^a = (p^0, \vec{p}))$ and a
spin of fields. Charges of fields will   
not be pointed out. The Weyl equation for massless fields
\begin{equation}
2 \vec{S}.
\vec{p} =   \Gamma p^0 
\label{weyl} 
\end{equation}
determines four states.
Here $S^{i} = \frac{1}{2}\varepsilon_{ijk} S^{jk},\;\; S^{ij} =
\frac{i}{4} \;[\gamma^i, \gamma^j],\;\;(i\in\{1,2,3\}) $, which
are the generators of the Lorentz transformations,
determine  the spin of states and $
\gamma^a, \;\;a \in\{0,1,2,3\} $  are the Dirac operators. 
The operator $\Gamma = \frac{-i}{3!} \varepsilon_{abcd}S^{ab}S_{cd}$ 
is one of the two  Casimir operators  of the
Lorentz group $SO(1,3)$ acting in the internal space of spins
only and defines the handedness of states. Two eigenstates of Eq.(\ref{weyl})
have left handedness ($< \Gamma > = r,\; r =
-1$), the other two have right handedness\footnote[1] { We shall make
use of the symbol $\Gamma$ for the operator and r (ro\v cnost in
slovenian language means handedness) for the corresponding 
eigenvalue. The symbol $h$ will be used 
for both, for the helicity operator and for its eigenvalue.}
($r = 1$). The left 
handed solutions have either left helicity 
\begin{equation}
 h = \frac{2 \vec{S}.
\vec{p}}{ |\vec{p^0}|},
 \label{h}
\end{equation}
( $h = -1$) and positive energy ($ p^0 = |p^0| $) or right
helicity ($h = 1$) and negative energy ($ p^0 = -|p^0|$). The
right handed solutions have either  right helicity ($ h = 1$)
and positive energy or left helicity ($h = -1$) and negative
energy. We shall denote the positive energy solutions by a
symbol $u$ and the negative enery solutions by a symbol $v$.
To determine the positive energy solution completly is enough to
tell the momentum $\vec{p}$, ( $p^0 =
|\vec{p}|$) and 
either handedness or helicity: $ u_{\vec{p}, L}
\equiv u_{\vec{p}, h = -1},$  $\; u_{\vec{p}, R}
\equiv u_{\vec{p}, h = 1},$ L and R stand for left and right
handedness, respectively. Equivalently it follows for the
negative energy solution
: $ v_{\vec{p}, L}
\equiv v_{\vec{p}, h = 1},$  $\; v_{\vec{p}, R}
\equiv v_{\vec{p}, h = -1}.$
We shall point out either helicity (h) or handedness (r),
depending on what will be more convenient. 
\\

\noindent
After quantizing the field the creation operators are defined,
creating the 
negative energy particles: $ d_{\vec{p}, L}^{(0)+}
\equiv d^{(0)+}_{\vec{p}, h = 1},$   $
d^{(0)+}_{\vec{p}, R} 
\equiv d^{(0)+}_{\vec{p}, h = -1},$ and the
positive energy particles: $ b^{(0)+}_{\vec{p}, L}
\equiv b^{(0)+}_{\vec{p}, h = -1},$   $
b^{(0)+}_{\vec{p}, R} 
\equiv b^{(0)+}_{\vec{p}, h = 1}.$ 
The field operator can then according to ref.\cite{bjor} be written as:
\begin{equation}
 \psi(x) = \sum_{r = \pm1} \sum_{\vec{p}, p^{02} 
= {\vec{p}}^2} \frac {1}{\sqrt{(2 \pi)^3}} ( b^{(0)}{
}_{\vec{p}, r} u_{\vec{p}, r} e^{ -ipx} +
d^{(0)}{ }_{\vec{p}, r} v_{\vec{p}, r}
e^{ ipx} ). 
\label{psi} 
\end{equation}
\noindent
To simplify the discussions we discretize the momentum and replace
the integral  with the sum. 
Then  the energy operator $ H^{(0)} = \int
d^3 \vec{x} \psi^+ p_0 \psi $ can be written as
\begin{equation}
 H^{(0)} = \sum_{r = \pm1} \sum_{\vec{p}, p^{02} =
\vec{p}^2} 
|p^0|\; ( b^{(0)+}_{\vec{p}, r}
b^{(0)}_{\vec{p}, r} - d^{(0)+}{
}_{\vec{p}, r} 
d^{(0)}_{\vec{p}, r} ).
\label{h0}
\end{equation}
\noindent
If the "totally empty" vacuum state is denoted by $|0>$, then
the vacuum 
state occupied by massless particles up to $ \vec{p}
= 0$ is ( due to discrete values of momenta ) equal to 
\begin{equation}
|\phi_{(0)}> = \prod_{\vec{p}, r}
d^{(0)+}_{\vec{p}, r} |0>. 
\label{phi0}
\end{equation}
The energy of such a vacuum state is
accordingly $<\phi_{0}| H^{(0)}|\phi_{0} > = \sum_{\vec{p}, r} 
E_{\vec{p}, r}^{(0)}, $ with $ E_{\vec{p}, r}^{(0)} =
-|\vec{p}|, $ which is of course infinite.
Accordingly the particle state of momentum $\vec{\acute{p}},
\acute{p}^0 = |\vec{\acute{p}}|, $ and handedness
$r$, with the energy, which is for $\acute{p}^0$ larger than the
energy of the vacuum state, can be written as $ b^{(0)+}_{\vec{\acute{p}}}
|\phi_0>.$

\section{ The charge conjugation \label{cc}}

\noindent
The symmetry operation of charge conjugation is associated with
the interchange of particles and antiparticles.
Introducing the charge conjugation operator $C$, with the
properties $ C^2 = C, \; C^+ = C, \; C \gamma^{a*} C^{-1} = -\gamma^a $ ,
where $(^+)$ 
stays for hermitian conjugation and $(^*)$ for complex
conjugation, one finds the charge conjugated 
field $ \psi(x)^C $ to the field $\psi(x)$ as $ \psi(x)^C = C
\psi(x). $ One accordingly finds for the charge conjugating
operator ${ \cal C }$, which affects creation and anihilation
operators 
\begin{eqnarray} 
{\cal C}\;\; b^{(0)+}_{\vec{p}, h = -1} \;\;{\cal C}^{-1}
= -d^{(0)}_{-\vec{p}, h = 1}, & {\cal C}\;\;
b^{(0)+}_{\vec{p}, h = 1} \;\; {\cal C}^{-1} 
= d^{(0)}_{-\vec{p}, h = -1}, 
\nonumber
\\
 {\cal C} \;\; d^{(0)+}_{\vec{p}, h = 1} \;\; {\cal C}^{-1}
= -b^{(0)}_{-\vec{p}, h = -1}, & {\cal C}\;\; 
d^{(0)+}_{\vec{p}, h = -1} \;\; {\cal C}^{-1} 
= b^{(0)}_{-\vec{p}, h = 1}.
\label{ch}
\end{eqnarray} 
\noindent
According to section \ref{zm} the left handed column concerns
the charge 
conjugation of left handed particles while the right handed column
concerns the charge conjugation of right handed particles. 
One easily finds that the hamiltonian $ H^{(0)}$ is invariant
under the charge conjugation operation. The charge conjugation
operation on the vacuum state $\;\; |\phi
_{(0)}> = 
\prod_{\vec{p}, r = \pm 1}  d^{(0)+}_
{\vec{p}, r}  |0> \;\;$  
should let it be invariant, since we want it as the physical
vacuum state to be charge conjugation invariant. To achieve that we
are, however, then forced to let the "totally empty" vacuum state $\;
|0> \;$ transform under charge conjugation as
\begin{equation}
 { \cal C}|0>\; = \prod_{\vec{p},r}\;(
b^{0 \dagger}_{\vec{p},r} d^{0 \dagger}_{\vec{p},r})\;|0> \label{c0} 
\end{equation}
The charge conjugated operator $\; 
b^{(0)+}_{\vec{p}, L}$ ( which generates
on a vacuum $\; |\phi_{(0)}> \;$ a one particle positive energy state of
left helicity )
annihilates in the vacuum state
$ \; |\phi_{(0)}> \; $ a  negative energy particle state of opposite
momentum and helicity and therefore generates a hole, which
manifests as an antiparticle. Handedness stays unchanged.

\section{ The massive Dirac field theory \label{mdf}}

\noindent
We shall first treat the case of the massive Dirac field, for
which the procedure is well known and simpler
than in the massive Majorana case and from which we can learn
the procedure. The
mass term $ \; \int d^3 \vec{x} m_{D} \bar{\psi} \psi = \int d^3
\vec{x} \; m_{D} \; ( \bar{\psi_{L}} \psi_{R} + \bar{\psi_{R}}
\psi_{L} )  \;$ can be written, if using the 
expression for $\psi$ from Eq.(\ref{psi}),  as follows
\begin{equation}
 H^{(1D)} = m_{D} \; \int d^3 \vec{x} \; \bar{\psi} \psi
= \sum_{h = \pm1} 
\sum_{\vec{p}}  
H^{(1D)}_{\vec{p}, h},\;\;\; 
H^{(1D)}_{\vec{p}, h} = m_D \;(\; b^{(0)+}_
{\vec{p}, h} d^{(0)}_{\vec{p}, h}
+ d^{(0)+}_
{\vec{p}, h} b^{(0)}_{\vec{p}, h}\;).
\label{dm} 
\end{equation}
If we define
\begin{eqnarray}
 N_{\vec{p}, h} = h^{+}_{\vec{p}, h} + h^{-}_{\vec{p}, h},
\nonumber
\\
 h^{+}_{\vec{p}, h} = b^{(0)+}_
{\vec{p}, h} b^{(0)}_{\vec{p}, h}, \;\; h^{-}_{\vec{p}, h} = 
d^{(0)+}_{\vec{p}, h} d^{(0)}_
{\vec{p}, h}, 
\label{hod}
\end{eqnarray}
one easily finds that $ \; [N_{\vec{p}, h},
H^{(0)}_{\vec{\acute{p}}, \acute{h}}] = 0 = 
[N_{\vec{p}, h},
H^{(1D)}_{\vec{\acute{p}}, \acute{h}}] \;$. We see that the
interaction term $\; H^{(1D)}_{\vec{p}, h} \;$ does not mix massless
states of different helicity. The appropriate basic states,
which are eigenstates of the operator for number of particles of
definite helicity $ \; N_{\vec{p}, h} \;$ are accordingly defined
either with $\; b^{(0)+}_ 
{\vec{p}, h = 1} \;$ and $\; d^{(0)+}_{\vec{p}, h = 1} \;$ with $h = 1$
( but of right and left handedness, respectively ) or with $\;
b^{(0)+}_ 
{\vec{p}, h = -1} \;$ and $\; d^{(0)+}_{\vec{p}, h = -1} \;$ with $ h =
-1$ (but of left and right handedness, respectively). The first
two basic states have $\; < N_{\vec{p}, h = 1} > = 1 \; $ and 
$\; < N_{\vec{p}, h = -1} > = 0 \;$, while the second two basic
states have   $\; < N_{\vec{p}, h = 1} > = 0  \;$ and 
$\; < N_{\vec{p}, h = -1} > = 1 \;$. 

\noindent
Diagonalizing 
 $
\; H^{(D)}_{\vec{p}, h}  =  H^{(0)}_{\vec{p},
h} + H^{(1D)}_{\vec{p}, h} \;$
 within the two basic states of
definite helicity ( but not handedness ), one finds 
that
\begin{eqnarray}
b^{+}_{\vec{p}, h} = \alpha_{\vec{p}} \; b^{(0)+}_{\vec{p},
h} + \beta_{\vec{p}} \; d^{(0)+}_
{\vec{p}, h}, & p^0 = \;\;\;|p^0|,
\nonumber
 \\
d^{+}_{\vec{p}, h} = \alpha_{\vec{p}} \; d^{(0)+}_{\vec{p},
h} - \beta_{\vec{p}} \; b^{(0)+}_
{\vec{p}, h}, & p^0 = -|p^0|, 
\nonumber
\\
\alpha_{\vec{p}} = \sqrt{\frac{1}{2}(1 +
\frac{|\vec{p}|}{|p^0|})}, & \beta_{\vec{p}} = \sqrt{\frac{1}{2}(1
- \frac{|\vec{p}|}{|p^0|})}, 
\nonumber
\\ 
 \;\;|p^0| = \sqrt{\vec{p}^2 + 
m_{D}{ }^2 }. &   
\label{bdmd}
\end{eqnarray}
 The operator $ b^{+}_{\vec{p}, h} $ creates a massive
positive energy one
particle state  ($ p^0 = |p^0| $)  and $
d^{+}_{\vec{p}, h} $ creates a massive negative energy one
particle state  
($ p^0 = - |p^0|$), both states have momentum $\vec{p}$ and
helicity $h$. Both 
are eigenstates of the hamiltonian for a massive  
Dirac field 
\begin{equation} 
H^{(D)} = \sum_{{\vec{p}, h}}\;  H^{(D)}_{\vec{p}, h} = 
\sum_{{\vec{p}, h}} \; ( H^{(0)}_{\vec{p}, h} +
H^{(1D)}_{\vec{p}, h} ) = \sum_{{\vec{p}, h}} \;
|p^0|  ( b^{+}_{\vec{p}, h}
b_{\vec{p}, h} - d^{+}_{\vec{p}, h} d_{\vec{p}, h} ) 
 \label{hd} 
\end{equation}
of momentum $\vec{p}$ and  helicity $h$. 
The state of the Dirac sea of massive particles is now
\begin{equation} 
|\phi_{(D)}> = \prod_{\vec{p}, h = \pm1}
d^{+}_{\vec{p}, h} |0> =  \pi_{\alpha} \; e^{- \sum_{\vec{p}, h}  
\frac{\beta_{\vec{p}}}{\alpha_{\vec{p}}} \; b^{(0)+}_{\vec{p}, h}
d^{(0)}_{\vec{p}, h} } \; |\phi_{(0)} >, \; \pi_{\alpha} =
\prod_{\vec{\acute{p}}} \alpha_{\vec{\acute{p}}}. 
\label{phid}
\end{equation}
All states up to $p^0 = -m_D$ are occupied and due to Eq.(\ref{c0})
it follows that $\; {\cal C}
|\phi_{(D)}>\; = |\phi_{(D)}>.\; $
 The interaction term
$H^{(1D)}$ causes the superposition of positive and negative
energy massless states (Eq.(\ref{bdmd})). One sees that the vacuum state of
massive particles can be understood as a coherent state of
particle and antiparticle pairs on the massless vacuum state.
The energy of the vacuum state of massive Dirac particles
$\; <\phi_{(D)}| H^{(D)}|\phi_{(D)} > = \sum_{\vec{p}, h} \; E_{\vec{p},
h}^{(D)}, \;$ with $ \;
E_{\vec{p}, h}^{(D)} = 
-\sqrt{\vec{p}^2 + m_D{ }^2 } \;$,  which is  infinite.\\

\noindent
According to  Eq.(\ref{bdmd}), the creation and annihilation operators 
for massive fields go in the limit when $\; m_D
\longrightarrow 0 \;$ to the creation and annihilation operators
for the massless case. 
\\

\noindent
A one particle state of energy $\; |p^0| =
\sqrt{ \vec{p}^2 + m_D^2 } \;$
can be written as $\; b^{+}_{\vec{p}, h} \; |\phi_{(D)}>,\;$
with $\; b^{+}_{\vec{p}, h}\;$ defined in Eq.(\ref{bdmd}). Also
this state 
becomes in the limit $m_{D} = 0$ a massless Weyl one particle
state of positive energy $\; |\vec{p}|\;$ above the sea of massless
particles.  \\ 

\noindent
One easily finds that $H^{(1D)}$ is invariant under charge
conjugation and so is therefore also $H^{(D)}$. Taking into
account  Eqs.(\ref{ch}) it follows
\begin{eqnarray}
{\cal C} \; b^{+}_{\vec{p}, h = -1} \; {\cal C}^{-1}
= \;-d^{}_{-\vec{p}, h = 1}, & {\cal C} \;\;
b^{+}_{\vec{p}, h = 1} \;\; {\cal C}^{-1} \;
= d^{}_{-\vec{p}, h = -1},
\nonumber
 \\
 {\cal C}\; d^{+}_{\vec{p}, h = 1} \;\;\; {\cal C}^{-1}
= -b^{}_{-\vec{p}, h = -1}, & {\cal C} \;
d^{+}_{\vec{p}, h = -1} \; {\cal C}^{-1} 
= b^{}_{-\vec{p}, h = 1}.
\label{chd}
\end{eqnarray}

\noindent
In the limit $\; m_D \longrightarrow 0 \;$ Eqs. (\ref{chd}) coincide with
Eqs. (\ref{ch}).
The charge conjugation transforms the particle of a momentum
$\vec{p}$ and helicity $h$ into the hole in the Dirac
sea of the  momentum $-\vec{p}$ and helicity $-h$.

\section{ The  massive Majorana field theory \label{mmf}}

\noindent
The Majorana mass term with only left handed fields $ \; m_{ML}
\; \int d^3  \vec{x} \; (\bar{\psi}_L + \bar{\psi}_L{ }^C)
(\psi_{L} + \psi_{L}^C) \; $
can be written, if using the expression for $\psi$ from
Eq.(\ref{psi}), with the summation going over the left handed fields
only and if taking into account the definition of charge
conjugation from section \ref{cc},
as follows
\begin{equation}
H^{(1M)}_{L} = m_{ML} \; \int d^3 \vec{x} \; ( \bar{\psi}_L +
\bar{\psi}_L{ }^C ) (\psi_L + \psi_L{ }^C)
= \sum_{(\vec{p})^+}  
H^{(1M)}_{\vec{p}, L}, 
\nonumber
\end{equation}
\begin{equation}
H^{(1M)}_{\vec{p}, L} = m_{ML} \; ( b^{(0)+}_
{\vec{p}, h = -1} b^{(0)+}_{-\vec{p}, h = -1} +
b^{(0)}_{-\vec{p}, h = -1} b^{(0)}_{\vec{p}, h = -1} + 
d^{(0)+}_
{\vec{p}, h = 1} d^{(0)+}_{-\vec{p}, h = 1} +  d^{(0)}_
{-\vec{p}, h = 1} d^{(0)}_{\vec{p}, h = 1} ).
\label{mlm}
\end{equation} 
The symbol $\; \sum_{(\vec{p})^+} $ means that the sum runs over
$\vec{p} $ on such a way that $\vec{p}$ and $-\vec{p}$ is
counted only once. 
Comparing the Majorana interaction term $\;H^{(1M)}{ }_{\vec{p},
L} \;$ of
Eq.(\ref{mlm}) with 
the Dirac interaction term of Eq.(\ref{dm}), one sees that in both
cases momentum $\vec{p}$ is conserved as it should be. In Eq.
(\ref{mlm}) the two creation operators appear with opposite momentum,
while in Eq.(\ref{dm}) the creation and annihilation 
operators appear with the same momentum. Because of that we
could pay attention in the Dirac case to a momentum $\vec{p}$,
without connecting ${\vec{p}}$ with ${-\vec{p}}$, while  
in the Majorana case we have to treat ${\vec{p}}$ and
${-\vec{p}}$ at the same time.

\noindent
The Majorana  mass 
term of only right handed fields follows from the mass term of
only 
left handed fields of  Eq.(\ref{mlm}) if we
exchange $h = -1$ with $h = 1$ and $h = 1$ with $h = -1$.
We shall treat here the left handed fields only. The
corresponding expressions for the
massive
Majorana right handed fields can be obtained from the left
handed ones by the above mentioned exchange of helicities
of fields.
\\

\noindent
It is easy to check that the charge conjugation operator ${\cal
C}$ from Eq. (\ref{ch}) leaves the interaction term of Eq.(\ref{mlm})
unchanged. Accordingly also the hamiltonian
\begin{equation}
 H^{(M)}_{\vec{p}, L } = H^{(0)}_{\vec{p}, L } + 
H^{(1M)}_{\vec{p}, L }  
\label{hm}
\end{equation} 
is invariant under the charge conjugation: $[H^{(M)}_{\vec{p}, L },
{\cal C} ] = 0. $ \\

\noindent
As in the Dirac massive case, it is meaningful to use the
operators $ h^{+}_{\vec{p}, h = -1} = b^{(0)+}_{\vec{p}, h = -1}
b^{(0)}_ {\vec{p}, h = -1}  $ and $ h^{-}_{\vec{p}, h = 1}
= d^{(0)+}_{\vec{p}, h = 1} 
d^{(0)}_ {\vec{p}, h = 1},$ to
choose the appropriate basis within which we shall diagonalize
the hamiltonian of Eq.(\ref{hm}). One can check that the operators
\begin{equation}
h^{+}_{\vec{p}} = h^{+}_{\vec{p}, h = -1} - h^{+}_{-\vec{p}, h
= -1},\;\; h^{-}_{\vec{p}} = h^{-}_{\vec{p}, h = 1} -
h^{-}_{-\vec{p}, h = 1}, 
\label{hom}
\end{equation}
which count the momentum of states, commute with the hamiltonian
of
Eq.(\ref{hm}) 
\begin{equation}
 [ h^{\pm}_{\vec{p}}, H^{(M)}_{\vec{p},L} ] = 0. 
\label{hmch}
\end{equation}
Since basic states, appropriate to describe the vacuum
state, should have momentum equal to zero to guarantee the zero
momentum of the vacuum,  one looks for the basic states with
$\; < h^{\pm}_{\vec{p}} > = 0. \;$ One finds four such states
 \begin{eqnarray}
|1> = & b^{(0)+}_{\vec{p}, h = -1} b^{(0)+}_{-\vec{p}, h = -1}
|0>, 
\nonumber
\\
|2> = &  \frac{1}{\sqrt{2}} ( 1 + b^{(0)+}_{\vec{p}, h = -1}
b^{(0)+}_{-\vec{p}, h = -1}  d^{(0)+}_{\vec{p}, h = 1}
d^{(0)+}_{-\vec{p}, h = 1} ) |0>, 
\nonumber
\\
|3> = & d^{(0)+}_{\vec{p}, h = 1}
d^{(0)+}_{-\vec{p}, h = 1} |0>\\ \hline 
|4> = &  \frac{1}{\sqrt{2}} ( 1 - b^{(0)+}_{\vec{p}, h = -1}
b^{(0)+}_{-\vec{p}, h = -1}  d^{(0)+}_{\vec{p}, h = 1}
d^{(0)+}_{-\vec{p}, h = 1} ) |0>. 
\label{bsms}
\end{eqnarray} 

\noindent
One finds that the state $|4>$ is  the eigenstate of
the hamiltonian of Eq.(\ref{hm}) with the eigenvalue zero. 
The hamiltonian applied on first three basic states defines a
matrix 

$$
\left( \begin{array}{ccc}
2|\vec{p}| & \sqrt{2} m & 0 \\
\sqrt{2} m & 0 & \sqrt{2} m \\
0 & \sqrt{2} m  & -2|\vec{p}|
\end{array} \right). \label{mass} $$ 

\noindent
Diagonalizing this matrix one finds three vectors and three
eigenvalues. The only 
candidate for the vacuum state is the state 
$\beta_{\vec{p}}{ }^2 |1> + (-) \sqrt{2}\;\alpha_{\vec{p}} \beta_{\vec{p}}
|2> + \alpha_{\vec{p}}{ }^2 |3>, $ with $\alpha_{\vec{p}} $ and
$\beta_{\vec{p}} $ defined in Eq.(\ref{bdmd}),
 with the 
eigenvalue $-2\sqrt{ |\vec{p}|^2 + m_{ML}{ }^2}$ 
which corresponds to the vacuum state of the $-2\sqrt{
|\vec{p}|^2 + m_{D}{ }^2}$ energy in the 
Dirac massive case $d^{+}_{\vec{p}, h = 1}
d^{+}_ {-\vec{p}, h = 1}  |0>.$ 
The Majorana vacuum state is accordingly
\begin{equation}
 |\phi_{(ML)}> = \prod_{(\vec{p})^{+}} \; |\phi_{M\vec{p}, L}>
, \; |\phi_{M\vec{p}, L} > = ( \beta_{\vec{p}}{ }^2 |1> +
(-)\sqrt{2} \; \alpha_{\vec{p}} \beta_{\vec{p}} |2> + \alpha_{\vec{p}}{ }^2
|3> ), 
\label{phim}
\end{equation}
\noindent
again with $ \alpha_{\vec{p}}$ and $ \beta_{\vec{p}}$ defined in
Eq.(\ref{bdmd}).

\noindent
As in the case of the Dirac sea, the Majorana sea can also be
written as an exponential operator working on a massless vacuum
state 
\begin{eqnarray}
|\phi_{(ML)}> = \pi_{\alpha^{+}}^2\; e^{- \sum_{(\vec{p})^{+}}
\frac{\beta_{\vec{p}}}{\alpha_{\vec{p}}} \; b^{(0)+}{ }
_{\vec{p}, h= -1}\; b^{(0)+}{ }
_{-\vec{p}, h= -1}\; } \cdot
\nonumber
\\
 e^{- \sum_{\vec{p}}
\frac{\beta_{\vec{p}}}{\alpha_{\vec{p}}} \; d^{(0)}{ }
_{\vec{p}, h= 1}\; d^{\;\; (0)}{ }
_{-\vec{p}, h= 1}\; }\; |\phi_{(0)'}>, 
\;\; \pi_{\alpha^{+}}^2 =
\prod_{(\vec{p})^{+}} \alpha_{\vec{p}}^2, 
\nonumber
\end{eqnarray}
if the massless vacuum state of only left handed particles is
written as $ |\phi_{(0)'}> = \prod_{(\vec{p}, L)}\;d^{(0)^{+}}{ }
_{\vec{p}, h= 1}\; d^{(0)^{+}}{ }
_{-\vec{p}, h= 1}.\; \; $

\noindent 
Compared to the Dirac particle case it should be noted that 
 for the Majorana case we had to combine both $\vec{p}$ and $-\vec{p}$
when  constructing the ground state
(Eq.(\ref{phim})).
The energy of the vacuum state of the Majorana left handed
particles  is
$\; <\phi_{(ML)}| H^{(M)}|\phi_{(ML)} > = \sum_{(\vec{p})^+, L} \;
E_{\vec{p}, L}^{(ML)}, \;$ with $ \;
E_{\vec{p}, L}^{(ML)} = 
-2 \sqrt{\vec{p}^2 + m_{ML}{ }^2 } \;$, which is the energy of two
majorana particles of momentum $p^a = ( -|p^0|, \vec{p}) $ and $p^a
= ( -|p^0|, -\vec{p}) $,
respectively. The energy of the Majorana sea is again infinite.
In the limit $ m_{ML} \longrightarrow 0 $ the Majorana sea
becomes the sea of   Weyl particles of only left handedness. \\

Concerning charge conjugation we see that with the somewhat
complicated transformation of the "totally empty" vacuum 
(Eq.(\ref{c0})) the Majorana physical vacuum $\; |\phi_{ML}>\;$
is charge conjugation invariant
\begin{equation}
 {\cal C} |\phi_{ML}> = |\phi_{ML}>.
\label{cml}
\end{equation}

\noindent
We have further to construct the one particle Majorana state to
generate a physical fermion.
The one particle Majorana states can be constructed as  superpositions
of states with $\; < h^{(+)}{ }_{\vec{p}} + h^{(-)}{ }_{\vec{p}} > =
\pm 1 \;$. One finds four times two states which fulfil this
condition 
\begin{eqnarray}
|5>\; = &  b^{(0)+}_{\vec{p}, h = -1} |0>,
\nonumber
\\
|6>\; = & b^{(0)+}_{\vec{p}, h = -1} d^{(0)+}_{\vec{p}, h =
1} d^{(0)+}_{-\vec{p}, h = 1} |0>,
\nonumber
 \\
\hline
|7>\; = &  d^{(0)+}_{\vec{p}, h = 1} |0>,
\nonumber
\\
|8>\; = &  d^{(0)+}_{\vec{p}, h = 1} b^{(0)+}_{\vec{p}, h =
-1} b^{(0)+}_{-\vec{p}, h = -1} |0>,
\nonumber
 \\
\hline\hline
|9>\; = &  b^{(0)+}_{-\vec{p}, h = -1} |0>,
\nonumber
\\
|10> = & b^{(0)+}_{-\vec{p}, h = -1} d^{(0)+}_{\vec{p}, h =
1} d^{(0)+}_{-\vec{p}, h = 1} |0>,
\nonumber
\\
\hline
|5> = &  d^{(0)+}_{-\vec{p}, h = 1} |0>,
\nonumber
\\
|6> = &  d^{(0)+}_{-\vec{p}, h = -1} b^{(0)+}_{\vec{p}, h =
-1} b^{(0)+}_{-\vec{p}, h = -1} |0>, 
\label{bsmp}
\end{eqnarray} 
The first four states have a momentum $\vec{p}$ and the last four 
states a momentum $-\vec{p}$. The hamiltonian $
H^{(M)}_{\vec{p}, L} $ defines on these states the block
diagonal four 
two by two matrices. The candidates for the states describing a
one particle state of momentum $\vec{p}$ on a vacuum 
states $|0> $ are states with energy which is for $ p^0 =
\sqrt{\vec{p}^2 + m_{ML}^2 } $ higher than the vacuum state.
One finds the corresponding operators
\begin{eqnarray}
b^{+}_{\vec{p}, h = -1}  = & -\beta_{\vec{p}}
\; b^{(0)+}_{\vec{p}, h = -1} + \alpha_{\vec{p}} \; b^{(0)+}_{\vec{p}, h =
-1} d^{(0)+}_{\vec{p}, h = 1}  d^{(0)+}_{-\vec{p}, h = 1}, 
\nonumber
\\
d^{+}_{\vec{p}, h = 1}  = & -\alpha_{\vec{p}}
\; d^{(0)+}_{\vec{p}, h = 1} + \beta_{\vec{p}} \; d^{(0)+}_{\vec{p}, h =
1} b^{(0)+}_{\vec{p}, h = -1}  b^{(0)+}_{-\vec{p}, h = -1}, 
\nonumber
\\ 
\hline
b^{+}_{-\vec{p}, h = -1}  = & -\beta_{\vec{p}}
\; b^{(0)+}_{-\vec{p}, h = -1} + \alpha_{\vec{p}} \;
b^{(0)+}_{-\vec{p}, h = 
-1} d^{(0)+}_{\vec{p}, h = 1}  d^{(0)+}_{-\vec{p}, h = 1}, 
\nonumber
\\
d^{+}_{-\vec{p}, h = 1}  = & -\alpha_{\vec{p}}
\; d^{(0)+}_{-\vec{p}, h = 1} + \beta_{\vec{p}} \; d^{(0)+}_{-\vec{p}, h =
1} b^{(0)+}_{\vec{p}, h = -1}  b^{(0)+}_{-\vec{p}, h = -1}, 
\label{bsmpp}
\end{eqnarray}
which when applied on a "totally empty" vacuum state $|0> $
generates the one 
particle states of momentum $\vec{p}$ (the first two operators)
and $-\vec{p}$ (the second two operators), respectively.
\\

\noindent
We would prefere to know, as in the Dirac massive case, the one
particle operators which when being applied on a Majorana vacuum
state $|\phi_{(ML)}> $ generates a one particle Majorana state with
chosen momentum $\vec{p}$ and which commute with the charge
conjugate operator ${\cal C}$ defined in Eq.(\ref{ch}).

\noindent
Requiring 
$\; B^+{ }_{\vec{p}, h = -1} |\phi_{M \vec{p}, L} > =
b^+_{\vec{p}, h = -1} |0> $ one finds $ B^+{ }_{\vec{p}, h
= -1} = \alpha_{\vec{p}} \;  b^{(0)+ }_{\vec{p}, h = -1} + 
\beta_{\vec{p}} \;  b^{(0)}_{-\vec{p}, h = -1},\; $ with
$\;\alpha_{\vec{p}}\; $ and $\; \beta_{\vec{p}} \; $ defined in
Eq.(\ref{bdmd}). 

\noindent
Accordingly it follows 
from the requirement $\; D^+{ }_{\vec{p}, h = 1} |\phi_{M \vec{p},
L} > = d^{+}_{\vec{p}, h = 1} |0> \;$ that $\; D^+{ }_{\vec{p}, h
= 1} = \beta_{\vec{p}} \;  d^{(0)+ }_{\vec{p}, h = 1} + 
\alpha_{\vec{p}} \;  d^{(0) }_{-\vec{p}, h = 1}. \;$
Taking into account that $\; {\cal C} \; B^+{ }_{\vec{p}, h = -1}
{\cal C}^{-1} = -D^{+}_{-\vec{p}, h = 1}  $
we may  conclude that the two operators
\begin{equation}
{\cal B}^+{ }_{ \pm\vec{p}, h = -1} = \alpha_{\vec{p}} \;(
b^{(0)+ }_{\pm\vec{p}, h = -1} - d^{(0)}_{\mp\vec{p}, h = 1} ) -
\beta_{\vec{p}} \; ( d^{(0)+}_{\pm\vec{p}, h = 1} - b^{(0)}{
}_{\mp\vec{p}, h = -1} ), 
\label{tpmp}
\end{equation} 

\noindent
 operating on the
Majorana vacuum state $|\phi_{ML}>$ generates 
the one paricle 
Majorana states of momentum $\pm\vec{p}$. It can easily be
checked that the Majorana particle is its own antiparticle $\; {\cal
C} \; {\cal B}^+{ }_{ \pm\vec{p}, h = -1} \; {\cal C}^{-1} =
{\cal B}^+{ }_{ \pm \vec{p}, h = -1} \;$.\\

\noindent
In the limit when $\;m_{ML} \longrightarrow 0\;$, the operator ${\cal
B}^+{ }_{ 
\pm\vec{p}, h = -1} $ operating on a vacuum state  $\; |\phi_{(ML)}>\;$, 
which goes to the vacuum state of the massless case of only left
handed paricles, gives a state of a Majorana
massless particle: $ ( b^{(0)+
}{ }_{\pm\vec{p}, h = -1} - d^{(0)}_{\mp\vec{p}, h = 1} )$ $ 
  d^{(0)+}{ }_{\vec{p}, h = 1}  d^{(0)+}{
}_{-\vec{p}, h = 1} $ $ \prod_{\vec{\acute{p}}, \acute{p} \neq p
} \; d^{(0)+}{
}_{\vec{\acute{p}}} |0>. $\\

\noindent
 One can
accordingly find the operators for 
right handed Majorana 
particles. 

\section{Conclusions}

\noindent
We have learned that it is indeed possible to define the
Majorana sea in the way the Dirac sea is defined.  It stays,
however, as an open problem to
study how this presentation can be used to better
understand the properties of the Majorana particles.

\section*{Acknowledgement } This work was supported by Ministry of
Science and Technology of Slovenia. One of the authors (N.
Manko\v c Bor\v stnik) acknowledges fruitfull discussions on
this topic with 
Mitja Rosina, the other author (H. B.
Nielsen) would like to thank the funds CHRX-CT94-0621, INTAS
93-3316, INTAS-RFBR 95-0567.

\end{document}